\documentclass{nature}
\usepackage{amsmath}    
\usepackage{graphicx}   
\usepackage{color}
\usepackage[10pt]{moresize}
\usepackage{graphics,graphicx}
\usepackage{amsfonts}
\usepackage{amssymb}
\usepackage{amscd}
\usepackage{amsmath}
\usepackage{enumerate}
\usepackage{epsfig}
\usepackage{subfigure}

\begin{document}

\title{Quantum Interference Induced Photon Blockade in a Coupled Single Quantum Dot-Cavity System}
\author{Jing Tang$^{1,2}$, Weidong Geng$^{1, \ast}$, Xiulai Xu$^{2, \dag}$}
\maketitle
\begin{affiliations}
\item
Institute of Photo-electronic Thin Film Devices and Technology,
Nankai University, Tianjin 300071, P.~R.~China
\item
Beijing National Laboratory for Condensed Matter Physics, Institute
of Physics, Chinese Academy of
Sciences, Beijing 100190, P.~R.~China \\
 $^\ast$ To whom correspondence should be addressed.
E-mail: gengwd@nankai.edu.cn\\
$^\dag$ To whom correspondence should be addressed. E-mail:
xlxu@iphy.ac.cn
\end{affiliations}

\baselineskip24pt

\maketitle

\begin{abstract}
We propose an experimental scheme to implement a strong photon
blockade with a single quantum dot coupled to a nanocavity. The
photon blockade effect can be tremendously enhanced by driving the
cavity and the quantum dot simultaneously with two classical laser
fields. This enhancement of photon blockade is ascribed to the
quantum interference effect to avoid two-photon excitation of the
cavity field. Comparing with Jaynes-Cummings model, the second-order
correlation function at zero time delay $g^{(2)}(0)$ in our scheme
can be reduced by two orders of magnitude and the system sustains a
large intracavity photon number. A red (blue) cavity-light detuning
asymmetry for photon quantum statistics with bunching or
antibunching characteristics is also observed. The
photon blockade effect has a controllable flexibility by tuning the
relative phase between the two pumping laser fields and the Rabi
coupling strength between the quantum dot and the pumping field.
Moreover, the photon blockade scheme based on quantum interference
mechanism does not require a strong coupling strength between the
cavity and the quantum dot, even with the pure
dephasing of the system. This simple proposal provides an effective
way for potential applications in solid state quantum computation
and quantum information processing.
\end{abstract}

\section*{Introduction}

Quantum information science (QIS) has been investigated intensively
for their fascinating potential applications in quantum computation,
cryptography, and metrology \cite{Ekert96, Knill01, Duan01,
Scarani09, Brien09}. Among these applications, the realization of
distribution, storage, and processing of quantum information in
single-photon level \cite{Kimble 08, Fushman08, Englund 12,
Faraon08,He2013} are of great importance. Up to now, various
platforms for implementing controllable single photons have been
proposed, such as single atoms coupled with micro-cavity systems
\cite{McKeever04, Hijlkema07, Wilk 07, Dayan 08}, or single quantum
dots integrated with photonic crystal cavities \cite{Toishi09,
Majumdar12, Faraon11, Hennessy07,Ishida2013}, optical
fibers\cite{Xu07, Xu08}, and surface plasmons \cite{Chang07}.

A key point for single photon manipulation is to realize photon
blockade. Photon blockade means that a first photon blocks the
second photon transmission induced by the quantum anharmonicity
ladder of energy spectrum with strong nonlinear interaction between
single photons, corresponding to an orderly output of photons one by
one with strong photon antibunching \cite{Imamoglu97}. However, for
a solid-state nanocavity with an embedded single quantum dot (QD),
the strong coupling condition with $g/\kappa\gg1$ is hard to achieve
due to the challenges of current micro-fabrication techniques for
high-quality nanocavity\cite{Majumdar2012,Reinhard2012,Brossard10},
where $g$ is the QD-cavity coupling strength and $\kappa$ is the
cavity decay rate. To solve this problem, the photon blockade with
strong sub-Poissonian light statistics based on bimodal-cavity
scheme has been theoretically proposed\cite{Majumdar2012, Zhang14}.
Meanwhile, strong photon blockade can be obtained in photonic
molecules with modest Kerr-nonlinearity of the photon using two
coupled photonic cavities\cite{Liew10, Bamba2011, Majumdar121,Brossard13}. Unfortunately, the strong photon nonlinearity is very difficult to
achieve at single photon level in most systems\cite{Boyd92}, and the
intracavity photon number is also low in the strong photon blockade
regime.

In this paper, a novel scheme for generating a strong photon
blockade with a single QD coupled to a nanocavity is proposed.
Different from the Jaynes-Cummings (JC) model, our scheme requires
an additional laser field to directly pump the single QD
simultaneously. By utilizing the optimal quantum interference (QI)
conditions, the cavity field exhibits the strong sub-Poissonian
statistics and a red (blue) cavity-light detuning asymmetry, which
is beyond the well known blockade mechanism induced by strong photon
nonlinearity. More importantly, a large intracavity photon number
(cavity output) is achieved with optimized parameters in photon
blockade regime, even for a modest QD-cavity coupling strength. The
$g^{(2)}(0)$ can be as low as 0.004 with a coupling
strength of $g/\kappa=2$. Consequently, it avoids the fabrication
challenges for preparing nanocavities with high quality factors.
Thus the proposed scheme can be used to obtain an ideal single
photon source\cite{Xu04}, which is more feasible experimentally.

\section*{Results}
\subsection{Model and Hamiltonian.}
We consider an excitonic two-level system of a single QD inside a
nanocavity. As shown in Fig. 1(a), the single QD is
coupled to the single mode nanocavity along $x$ axis with a
cavity frequency $\omega_c$ and QD-cavity coupling strength $g$. The
nanocavity is driven by a weak laser field with frequency
$\omega_p$, coupling strength $\eta$ and cavity
decay rate $\kappa$. Additional pump field along $y$ axis is applied
to pump the single QD directly with a frequency of $\omega_L$, and
provides a Rabi coupling strength $\Omega$. Figure 1(b)
shows the level structure for a single QD. In particular, even
without cavity driven field, the excitonic $|g\rangle\leftrightarrow
|e\rangle$ transition with frequency $\omega_a$ remains coupled to
the nanocavtiy through the vacuum-stimulated Bragg scattering
induced by the pump field \cite{Esslinger10,Deng14,Mottl}.

Using rotating wave approximation, the QD-cavity Hamiltonian can be
described by$(\hbar=1)$
\begin{align}
\hat{H}&= \omega_c \hat{a}^\dag\hat{a} + \omega_a \hat{\sigma}_{ee}
+ g(\hat{a}^\dag \hat{\sigma}_{ge} + \hat{a}
\hat{\sigma}_{eg}) \nonumber \\
&+\eta(\hat{a} e^{i\omega_pt} + \hat{a}^\dag
e^{-i\omega_pt})\nonumber \\
&+\Omega(e^{i\omega_L t + i\theta}\hat{\sigma}_{ge} +e^{-i\omega_L t
-i\theta}\hat{\sigma}_{eg}),
\end{align}%
where $\hat{a}$ and $\hat{a}^\dag$ are the cavity mode annihilation
and creation operators, $\hat{\sigma}_{ij}=|i\rangle\langle j|$ are
the QD spin projection operators with $i, j = {e,g}$ labeling the
two involved levels, and $\theta$ is the relative phase between QD
pumping field and cavity driven field.

For simplicity, we assume $\omega_p=\omega_L$ and $\omega_c=\omega_a$.
In the rotating frame with laser frequency $\omega_p$ by utilizing the unitary transformation $U$,
\begin{align}
U(t) & = {\rm exp}(-i \omega_p \hat{a}^\dag \hat{a}t -i\omega_p
\hat{\sigma}_{ee}t),
\end{align}
the interaction Hamiltonian of the QD-cavity system will be time-independent and can be rewritten as
\begin{align}
\hat {H}_{I}  &= U^\dag \hat{H}U -i
U^\dag\frac{\partial}{\partial t} U \nonumber\\
& = \Delta_c \hat{a}^\dag \hat{a} + \Delta_c \hat{\sigma}_{ee}+
g(\hat{a}^\dag \hat{\sigma}_{ge} + \hat{a} \hat{\sigma}_{eg})  \nonumber \\
&+ \eta(\hat{a} + \hat{a}^\dag)+ \Omega(e^{i\theta}\hat{\sigma}_{ge}
+e^{-i\theta} \hat{\sigma}_{eg}), \label{interaction}
\end{align}
where $\Delta_c = \omega_c - \omega_p= \omega_a - \omega_L$ is the
cavity-light detuning. Similar to JC model, the new QI model with
the Hamiltonian in Eq.~(\ref{interaction}) has an additional pump
laser coupling the single QD directly. As discussed below, the
relative phase $\theta$ of the two laser fields plays a significant
role in the photon blockade effect.

Without the pump field for $\Omega =0$, the Hamiltonian in
Eq.~(\ref{interaction}) is transformed to JC model. Neglecting the
effect of weak driven field, the Hamiltonian can be exactly solved
by projecting to a closed subspace with eigenstate basis
$|n,g\rangle$ and $|n-1,e\rangle$, where $n$ is the number of photon
excitation. Figure 1(c) shows the anharmonicity ladder
of energy spectrum of JC model \cite{Birnbaum 05}, in which the
dressed state $|n,+(-)\rangle$ represents the higher (lower) energy
level of the $n$-th excited states with energy eigenvalues $E_{n\pm}
= n\Delta_c \pm g\sqrt{n}$, where $g\sqrt{n}$ is the vacuum-Rabi
splitting of the $n$-th excited states. When the first excited
states are resonant with the laser field ($\Delta_c = \pm g$), the
energy levels of the second energy eigenstate $|2,\pm\rangle$ are
off-resonance with an energy gap of $\Delta' = (2-\sqrt{2})g$. In
strong coupling limit $g\gg \kappa$, the process of two-photon
excitation is strongly suppressed and photon blockade effect is
enhanced with $g^{(2)}(0)\sim 0$. It means a first photon ``blocks''
the second photon transmission to the cavity due to the far
off-resonance two-photon absorption, where the second-order
correlation function $g^{(2)}(0)={<\hat{a}^\dagger\hat{a}^\dagger
\hat{a}\hat{a}>}/{<\hat{a}^\dagger\hat{a}
>^2}$ describes the quantum statistics of the photon field.

\subsection{Quantum interference mechanism.} The photon blockade with anharmonic JC ladder is only achievable in
a strong coupling regime, which is difficult to obtain in a single
QD-nanocavity system. In our scheme, beyond the above photon
blockade mechanism of anharmonic ladder with $g\gg \kappa$, the
strong photon blockade can be achieved even at a moderate QD-cavity
coupling regime by simultaneously driving the cavity field and
pumping the single QD as illustrated in Fig. 1(a). Since the applied
pumping and driving fields are weak, the energy spectrum should be
almost same with JC model as shown in Fig. 1(c). Because of the
non-conserved excitation numbers, we cannot build a closed subspace
with the $n$th block spanned by $|n-1,e\rangle$ and $|n,g\rangle$.
As a result, the Hamiltonian matrix can not be diagonalized exactly
in the closed subspace. However, it can be diagonalized in the
subspaces defined by a given excitation number of the cavity field.
To understand the origin of the strong photon blockade, the
wavefunction can be written as ~\cite{Bamba2011}
\begin{align}
|\psi\rangle &= \sum_{n=0}^{\infty} C_{n,g}|n,g\rangle
+\sum_{n=1}^{\infty} C_{n-1,e}|n-1,e\rangle.
\end{align}
$|C_{n,g}|^2$ and $|C_{n-1,e}|^2$ represent the probabilities
of eigenstates $|n,g\rangle$ and $|n-1,e\rangle$, respectively. For
the photon blockade case, we just need to cut off the photons into the
two-photon excitation subspace with $n=2$. So the wave function for
the system can be expanded as: $|\psi\rangle =
C_{0,g}|0,g\rangle + C_{1,g}|1,g\rangle + C_{0,e}|0,e\rangle + C_{1,e}|1,e\rangle
+ C_{2,g}|2,g\rangle.$ To obtain the steady state solution,
these probability coefficients are satisfying
\begin{subequations}\label{stead-equation}
\begin{align}
i\dot{C}_{0,g} = &\eta C_{1,g} + \Omega e^{i\theta} C_{0,e} = 0, \\
i\dot{C}_{2,g} = &\sqrt{2}\eta C_{1,g} + (2\Delta_c -i2\kappa)C_{2,g} +\sqrt{2}g C_{1,e} = 0, \\
i\dot{C}_{1,e} = &\Omega e^{-i\theta} C_{1,g} + \eta C_{0,e} +
\sqrt{2}g C_{2,g} + (2\Delta_c -i\kappa-i\gamma) C_{1,e} = 0,
\end{align}
\end{subequations}
To suppress the two photon excitation, a condition $C_{2,g}=0$ is
required. In this limit, all higher photon excitations with $n\geq2$
are eliminated, resulting in only one excited photon in the
nanocavity. It should be noted that this blockade mechanism is
different from the strong coupling mechanism, where the higher
photon excitations are far off-resonance due to anharmonicity of
energy spectrum. The mechanism with strong photon blockade
is ascribed to the quantum interference effect with different transition paths
as shown in Fig. 1(d). Following the transition from $|0,g\rangle$ to $|1,g\rangle$ excited by the driven field, the
interference can happen between the two paths, the direct transition
$|1,g\rangle$ $\xrightarrow{\sqrt{2} \eta}$ $|2,g\rangle$ and the
transition $|1,g\rangle$ $\xrightarrow{\Omega e^{i\theta}}$
$|1,e\rangle$ $\xrightarrow{\sqrt{2} g}$ $|2,g\rangle$. In absence
of the pump field with $\Omega=0$ , we can not find any non-trivial solution from Eqs. (\ref{stead-equation}). When the pump field is
applied, from the Eqs. (\ref{stead-equation}) we can obtain the
stead solution satisfying
\begin{align}\label{con-equation}
C_{1,g} &= -\frac{\Omega e^{i\theta}}{\eta}C_{0,e}, \nonumber \\
C_{1,g} &= -\frac{g}{\eta}C_{1,e},\nonumber \\
C_{1,g} &= -\frac{(2\Delta_c -i\kappa -i\gamma)\Omega
e^{i\theta}}{\Omega^2 - \eta^2}C_{1,e}.
\end{align}

In vacuum-Rabi splitting with light-cavity detuning $\Delta_c = \pm
g$, the intracavity photon number should be large when the single photon
level is excited resonantly. By solving Eq.~(\ref{con-equation})
with $\Delta_c = \pm g$, the optimized relative phase $\theta_{\text
opt}$ and Rabi coupling strength $\Omega_{\text opt}$ are given by:
\begin{align} \label{OEQ}
&\theta_{\text opt} =
\begin{cases}
{\text arctan}(\frac{\kappa+\gamma}{2g}),~~~ \theta\in [0,
\frac{\pi}{2}], ~~~{\text{for}}~~~ \Delta_c =g \nonumber \\
{\text arctan}(-\frac{\kappa+\gamma}{2g}), ~~~\theta\in
[\frac{\pi}{2}, \pi] , ~~~{\text{for}}~~~ \Delta_c = -g \nonumber\\
\end{cases}, \nonumber \\
&\Omega_{\text opt} = {\eta}\frac{\mathcal {R} + \sqrt{\mathcal
{R}^2 +4}}{2},
\end{align}
where ${\mathcal R} =
\sqrt{4+\left(\frac{\kappa+\gamma}{g}\right)^2}$. The optimal
parameter $\theta_{\text opt}$ is dependent on the sign of
$\Delta_c$, which means that photon blockade relies on the specific
laser frequency. The optimal QI conditions in Eq~(\ref{OEQ}) are the
main results of this work.

Generally, when a cavity photon field has photon blockade effect,
multiple photon excitations ($n\geq2$) are suppressed by strong
photon nonlinearity or strong exciton-photon coupling. Here,
however, the photon blockade can be realized with completely
eliminating the two-photon excited states with $g^{(2)}(0)\sim 0$ by
using the QI mechanism with optimized conditions of Eq.~(\ref{OEQ}),
even for a moderate exciton-photon coupling strength. In the rest of
the paper, we take nanocavity decay rate $\kappa/2\pi = 20$ GHz,
single quantum dot spontaneous decay rate $\gamma/2\pi = 1.0$ GHz,
and weak cavity driven strength $\eta=0.1\kappa$.

\subsection{Numerical simulation.} By solving the time dependent master equation (see Methods), the
second-order correlation function $g^{(2)}(0)$ was calculated with
(without) the laser for pumping the quantum dot in QI (JC) model.
Figure 2(a) shows the minimum values of $g^{(2)}(0)$ for JC model
with $\Omega=0$, and of $g^{(2)}(0)$ for QI model with
$(\Omega,\theta)=(\Omega_{\text{opt}},\theta_{\text{opt}})$ as a
function of QD-cavity coupling strength $g$. Similar to the JC
model, the second-order correlation function $g^{(2)}(0)$
monotonically decreases with increasing the coupling strength $g$,
which suppresses the two-photon excitation due to a gradual increase
of two-photon absorption energy gap $\Delta'$. Surprisingly, photon
blockade effect in the QI model is tremendously enhanced comparing
with JC model at a specified coupling strength. For example, when
log$_{10}g^{(2)}(0)= -1.715$ (as shown with the black-dashed line in
Fig. 2(a)), the required coupling strength for JC model is $g/\kappa
= 12$ while that for QI model is only 1.01. This indicates that a
strong photon blockade can be achieved in a relative weak coupling
strength in QI model.

It can be seen that the $g^{(2)}(0)$ of JC model quasi-linearly
decreases as a function of coupling strength $g$. However, the
$g^{(2)}(0)$ of QI model drops much more quickly, indicating that
the $g^{(2)}(0)$ of QI model is more sensitive to $g$. To clearly
show the difference between the two models, the ratio $\Gamma =
g^{(2)}_{\text{JC}}(0)/g^{(2)}_{\text{QI}}(0)$ as a function of
coupling strength $g$ is plotted in Fig.~2(b). At a strong coupling
regime or even a moderate regime with $g/\kappa>2$, the photon
blockade for QI model is enhanced by two orders of magnitude. It
shows that a strong photon anti-buntuning ($g^{(2)}(0)\ll 1$) with
sub-Poissonian quantum statistics for cavity field output can be
easily achieved using the quantum interference method.

With a moderate QD-cavity coupling strength $g=2\kappa$, the
second-order correlation function $g^{(2)}(0)$ and the intracavity
photon number ${n}_c$ as a function of cavity-light detuning
$\Delta_c$ are shown in Fig.~3(a) and 3(b), respectively. For JC
model, both $g^{(2)}(0)$ and ${n}_c$ are symmetric for the red and
blue detuning with $\Delta=\pm g$, and are phase-independent for the
weak cavity driven field. However, asymmetric structures of QI model
for both $g^{(2)}(0)$ and ${n}_c$ are observed. With an optimized
phase condition $\tan\theta=(\kappa+\gamma)/(2g) = 0.2625$ and a red
detuning $\Delta_c=g$, photon blockade can be observed at the
position of $\Delta_c\approx g$ with $g^{(2)}(0)\approx 0$,
indicating sub-Poissonian quantum statistics for cavity field. But
at the position with $\Delta_c\approx -g$, the $g^{(2)}(0)$ is close
to unity, which is similar to the results in JC model (black solid
line). A reversed result can be obtained with the optimized phase
condition at blue detuning $\Delta_c=-g$, as shown by the
dash-dotted red line in Fig.~3(a).

The minimum $g^{(2)}(0)$ is about 0.004 when the laser field is
tuned to satisfy the optimized QI conditions of Eq.~(\ref{OEQ}).
Especially, even in the photon blockade regime, an intracavity
photon number $n_c$ of about 0.06 is still larger than the maximum
intracavity number in the JC model [as shown in Fig.~3(b)], which is
a key factor for ideal single photon sources with a large cavity
output~\cite{Birnbaum 05}. Figure 3(c) shows $g^{(2)}(\tau)$ as a
function of time. Anti-bunching effect with $g^{(2)}(0)<1$ and
$g^{(2)}(0)<g^{(2)}(\tau)$ indicates the output light is
sub-Poissonian and antibunching\cite{Hennrich05}. The
$g^{(2)}(\tau)$ rises to unity at a time $\tau\simeq 50$ ps, which
is consistent with the lifetime $\tau=1/(\gamma+\kappa) =48$ ps for
the dressed states $|1,\pm\rangle$ \cite{Birnbaum 05, Reinhard2012}.

To further investigate the red-blue detuning asymmetry, we
calculated $g^{(2)}(0)$ as a function of cavity-light detuning
$\Delta_c$ and relative phase $\theta$ with an optimized Rabi
coupling strength $\Omega/g=0.124$. As illustrated in Fig.~4, a red
(blue) detuning asymmetric feature for $g^{(2)}(0)$ is observed,
which is strongly correlated to the relative phase $\theta$. For
example, with a phase of $\theta/\pi \approx 0.08,~ \tan\theta>0$,
the $g^{(2)}(0)$ at a red detuning position of $\Delta_c\approx g$
approaches its minimum, which exhibits a strong sub-Poissonian
quantum statistics, whereas at the blue detuning the $g^{(2)}(0)$ is
close to unity. Similar features can be observed with phases at
$-0.082\pi\pm\pi$ for the blue detuning case with $\Delta_c\approx
-g$. Therefore in the QI model, the relative phase $\theta$ is
non-trivial and significantly influences the cavity quantum
statistics and output, which can not be eliminated by gauge
transformation. The simulation results by solving master equation
verify the theoretical prediction with optimized QI conditions in
Eq.~(\ref{OEQ}).

Figure~5(a) and 5(b) show the contour plots of $g^{(2)}(0)$ and
${n}_c$ as a function of $\Delta_c$ and $\Omega$ with a fixed phase
$\theta_{\rm opt}/\pi=0.082$. As expected, a strong photon blockade
should occur near the red detuning with $\Delta_c \approx g$. While
for blue detuning with $\Delta_c \approx -g$, there is no strong
blockade because the phase of 0.082$\pi$ is not an optimized value
in this case. Therefore, a higher intracavity photon number for blue
detuning regime is expected as shown in Fig.~5(b). Note that at red
detuning with $\Delta_c \approx g$, intracavity photon number $n_c$
is still much larger than the mean photon number ${n}_c =
(\eta/\kappa)^2=0.01$ in an empty cavity at strong photon blockade
regime. This means that this scheme can achieve an ideal single
photon source using solid-state single quantum dots with a strong
photon blockade and a large cavity output. In fact, a moderate
QD-cavity coupling strength $g$ is sufficient for this purpose,
which means that we do not need high quality factors (${Q}$) for the
nanocavities. In addition, the calculations show that photon
blockade effect can survive with a relatively large parameter
variation. As a result, the robustness of photon blockade for single
QDs does not need to perfectly satisfy the optimal QI conditions in
Eq.~(\ref{OEQ}), which should be more easily to be achieved
experimentally. In certain regimes, $g^{(2)}(0)$ with strong
super-Poissonian quantum statistics is also observed for
off-resonant excitation at $\Delta_c \approx 0.16g$ but with an
ultra-low intracavity photon number $n_c=1.613\times10^{-5}$.

So far, we did not consider the effect of pure dephasing, which could affect the polarization~\cite{Majumdar2011},
linewidth~\cite{Majumdar2010,Yamaguchi2008}, photon
statistics and cavity transmission~\cite{Majumdar12,Englund2012,Englund10,Auffeves2010} in solid state QD-cavity systems. Next, we study the effect of pure dephasing
$\gamma_d$ on photon blockade in the QI model by adding a
Lindblad term $\frac{\gamma_d}{2}\mathcal
{\cal{D}}[\hat{\sigma}_{eg}\hat{\sigma}_{ge}]\rho$ in the master
equation. Figure~6 shows the second-order
correlation function $g^{(2)}(0)$ and the intracavity photon number
$n_c$ with different pure dephasing rates. It can be seen that the $g^{(2)}(0)$ and $n_c$ still maintain the red-blue
detuning asymmetry. With increasing the pure dephasing rate,
$n_c$ does not change too much, but $g^{(2)}(0)$
increases near the red detuning with $\Delta_c \approx g$, while
remains the same at the blue detuning with $\Delta_c \approx -g$.
Nevertheless, the qualitative nature of the photon blockade is unchanged. For a typical pure dephasing rate
$\gamma_d=0.5\gamma$~\cite{Majumdar2011}, the $g^{(2)}(0)$ at the
red detuning with $\Delta_c \approx g$ is 0.01, and the
corresponding $n_c$ is still large for a coupling strength
$g=2\kappa$, which still can be treated as an
ideal single photon source with photon blockade.

\section*{Discussion} We proposed a new QI model with a
simple configuration by simultaneously driving the cavity field and
the single QD and realized strong photon blockade in a QD-cavity
system. Photon distributions with strongly antibunching effect and
sub-Poissonian statistics have been observed by numerically solving
the master equation using optimized phase $\theta$ and the coupling
strength $\Omega$. Furthermore, a red (blue) detuning asymmetry for
photon blockade has been observed. Photon blockade with a large
intracavity number for quantum dot shows a strong robustness, which
can be easily realized experimentally with considering the pure dephasing. From a practical point of view, it might be not easy to excite quantum dot and cavity separately. However, several schemes have been demonstrated successfully by using different pumping pulse widths \cite{Ishida2013}, or by spatially/spectrally decoupling the driving fields for quantum dot and cavity \cite{Majumdar2013,Flagg2009,Ates2009}. We believe the proposed scheme with QI mechanism could be very helpful for applications in various cavity
quantum-electrodynamics systems.

\section*{Methods} In order to demonstrate the photon
blockade, we investigated the quantum statistics of the nanocavity
field by solving quantum master equation numerically. Considering
the dissipation of the cavity with decay rate $\kappa$ and QD
spontaneous emission rate $\gamma$, without the nonradiative pure
dephasing, the master equation of the dynamics of single QD-cavity
system satisfies,
\begin{equation}\label{master equation}%
{\cal{L}}\rho = -i [\hat{H}_{I}, {\rho}] + \frac{\kappa}{2} \mathcal
{\cal{D}}[\hat{a}]\rho + \frac{\gamma}{2} \mathcal
{\cal{D}}[\hat{\sigma}_{ge}]\rho ,
\end{equation}
where $\rho$ is density matrix of QD-cavity system, $\hat{H}_{I}$ is
the time-independent interaction Hamiltonian of Eq.~(3), ${\cal{L}}$
is Liouvillian superoperator, and $\mathcal {D}[\hat{o}]\rho=
2\hat{o} {\rho} \hat{o}^\dag - \hat{o}^\dag \hat{o}{\rho} - {\rho}
\hat{o}^\dag \hat{o}$ is the Lindblad type of dissipation.
 Then the steady state
intracavity photon number ${n}_c = {\rm Tr}(\hat{a}^\dag
\hat{a}\rho_s)$ and the second-order correlation function
$g^{(2)}(0)={\rm Tr}(\hat{a}^\dagger\hat{a}^\dagger
\hat{a}\hat{a}\rho_s)/{n}^2_c$ can be obtained by calculating the
steady state density matrix with ${\cal L}\rho_s = 0$ using
Quantum Optics Toolbox \cite{Tan99}.

\noindent\textbf{Acknowledgments}\\
This work was supported by the National Basic Research Program of
China under Grant No. 2013CB328706 and 2014CB921003; the National
Natural Science Foundation of China under Grant No. 91436101,
11174356 and 61275060; the Strategic Priority Research Program of
the Chinese Academy of Sciences under Grant No. XDB07030200; and the
Hundred Talents Program of the Chinese Academy of Sciences. We thank
Y. Deng for very helpful discussions.

\noindent\textbf{Author Contributions}\\
J.T. performed calculations. X.X. and W.G. supervised the project. J.T. and X.X. wrote the paper and all authors reviewed the manuscript.

\noindent\textbf{Additional Information}\\
The authors declare no competing financial interests.

\bibliographystyle{naturemag}

\begin{thebibliography}{10}
\expandafter\ifx\csname url\endcsname\relax
  \def\url#1{\texttt{#1}}\fi
\expandafter\ifx\csname urlprefix\endcsname\relax\def\urlprefix{URL }\fi
\providecommand{\bibinfo}[2]{#2}
\providecommand{\eprint}[2][]{\url{#2}}

\bibitem{Ekert96}
\bibinfo{author}{Ekert, A.} \& \bibinfo{author}{Jozsa, R.}
\newblock \bibinfo{title}{Quantum computation and shor's factoring algorithm}.
\newblock \emph{\bibinfo{journal}{Rev. Mod. Phys.}}
  \textbf{\bibinfo{volume}{68}}, \bibinfo{pages}{733--753}
  (\bibinfo{year}{1996}).

\bibitem{Knill01}
\bibinfo{author}{Knill, E.}, \bibinfo{author}{Laflamme, R.} \&
  \bibinfo{author}{Milburn, G.~J.}
\newblock \bibinfo{title}{A scheme for efficient quantum computation with
  linear optics}.
\newblock \emph{\bibinfo{journal}{Nature}} \textbf{\bibinfo{volume}{409}},
  \bibinfo{pages}{46--52} (\bibinfo{year}{2001}).

\bibitem{Duan01}
\bibinfo{author}{Duan, L.-M.}, \bibinfo{author}{Lukin, M.~D.},
  \bibinfo{author}{Cirac, J.~I.} \& \bibinfo{author}{Zoller, P.}
\newblock \bibinfo{title}{Long-distance quantum communication with atomic
  ensembles and linear optics}.
\newblock \emph{\bibinfo{journal}{Nature}} \textbf{\bibinfo{volume}{414}},
  \bibinfo{pages}{413--418} (\bibinfo{year}{2001}).

\bibitem{Scarani09}
\bibinfo{author}{Scarani, V.} \emph{et~al.}
\newblock \bibinfo{title}{The security of practical quantum key distribution}.
\newblock \emph{\bibinfo{journal}{Rev. Mod. Phys.}}
  \textbf{\bibinfo{volume}{81}}, \bibinfo{pages}{1301--1350}
  (\bibinfo{year}{2009}).

\bibitem{Brien09}
\bibinfo{author}{O'Brien, J.~L.}, \bibinfo{author}{Furusawa, A.} \&
  \bibinfo{author}{Vu\v{c}kovi\'{c}, J.}
\newblock \bibinfo{title}{Photonic quantum technologies}.
\newblock \emph{\bibinfo{journal}{Nat. Photon.}} \textbf{\bibinfo{volume}{3}},
  \bibinfo{pages}{687--695} (\bibinfo{year}{2009}).

\bibitem{Kimble08}
\bibinfo{author}{Kimble, H.~J.}
\newblock \bibinfo{title}{The quantum internet}.
\newblock \emph{\bibinfo{journal}{Nature}} \textbf{\bibinfo{volume}{453}},
  \bibinfo{pages}{1023--1030} (\bibinfo{year}{2008}).

\bibitem{Fushman08}
\bibinfo{author}{Fushman, I.} \emph{et~al.}
\newblock \bibinfo{title}{Controlled phase shifts with a single quantum dot}.
\newblock \emph{\bibinfo{journal}{Science}} \textbf{\bibinfo{volume}{320}},
  \bibinfo{pages}{769--772} (\bibinfo{year}{2008}).

\bibitem{Englund12}
\bibinfo{author}{Englund, D.} \emph{et~al.}
\newblock \bibinfo{title}{Ultrafast photon-photon interaction in a strongly
  coupled quantum dot-cavity system}.
\newblock \emph{\bibinfo{journal}{Phys. Rev. Lett.}}
  \textbf{\bibinfo{volume}{108}}, \bibinfo{pages}{093604}
  (\bibinfo{year}{2012}).

\bibitem{Faraon08}
\bibinfo{author}{Faraon, A.} \emph{et~al.}
\newblock \bibinfo{title}{Coherent generation of non-classical light on a chip
  via photon-induced tunnelling and blockade}.
\newblock \emph{\bibinfo{journal}{Nat. Phys.}} \textbf{\bibinfo{volume}{4}},
  \bibinfo{pages}{859--863} (\bibinfo{year}{2008}).

\bibitem{He2013}
\bibinfo{author}{He, Y.-M.} \emph{et~al.}
\newblock \bibinfo{title}{On-demand semiconductor single-photon source with
  near-unity indistinguishability}.
\newblock \emph{\bibinfo{journal}{Nat. Nanotechnol.}}
  \textbf{\bibinfo{volume}{8}}, \bibinfo{pages}{213--217}
  (\bibinfo{year}{2013}).

\bibitem{McKeever04}
\bibinfo{author}{McKeever, J.} \emph{et~al.}
\newblock \bibinfo{title}{Deterministic generation of single photons from one
  atom trapped in a cavity}.
\newblock \emph{\bibinfo{journal}{Science}} \textbf{\bibinfo{volume}{303}},
  \bibinfo{pages}{1992--1994} (\bibinfo{year}{2004}).

\bibitem{Hijlkema07}
\bibinfo{author}{Hijlkema, M.} \emph{et~al.}
\newblock \bibinfo{title}{A single-photon server with just one atom}.
\newblock \emph{\bibinfo{journal}{Nat. Phys.}} \textbf{\bibinfo{volume}{3}},
  \bibinfo{pages}{253--255} (\bibinfo{year}{2007}).

\bibitem{Wilk07}
\bibinfo{author}{Wilk, T.}, \bibinfo{author}{Webster, S.~C.},
  \bibinfo{author}{Kuhn, A.} \& \bibinfo{author}{Rempe, G.}
\newblock \bibinfo{title}{Single-atom single-photon quantum interface}.
\newblock \emph{\bibinfo{journal}{Science}} \textbf{\bibinfo{volume}{317}},
  \bibinfo{pages}{488--490} (\bibinfo{year}{2007}).

\bibitem{Dayan08}
\bibinfo{author}{Dayan, B.} \emph{et~al.}
\newblock \bibinfo{title}{A photon turnstile dynamically regulated by one
  atom}.
\newblock \emph{\bibinfo{journal}{Science}} \textbf{\bibinfo{volume}{319}},
  \bibinfo{pages}{1062--1065} (\bibinfo{year}{2008}).

\bibitem{Toishi09}
\bibinfo{author}{Toishi, M.}, \bibinfo{author}{Englund, D.},
  \bibinfo{author}{Faraon, A.} \& \bibinfo{author}{Vu\v{c}kovi\'{c}, J.}
\newblock \bibinfo{title}{High-brightness single photon source from a quantum
  dot in a directional-emission nanocavity}.
\newblock \emph{\bibinfo{journal}{Opt. Express.}}
  \textbf{\bibinfo{volume}{17}}, \bibinfo{pages}{14618--14626}
  (\bibinfo{year}{2009}).

\bibitem{Majumdar12}
\bibinfo{author}{Majumdar, A.}, \bibinfo{author}{Bajcsy, M.} \&
  \bibinfo{author}{Vu\v{c}kovi\'{c}, J.}
\newblock \bibinfo{title}{Probing the ladder of dressed states and nonclassical
  light generation in quantum-dot-cavity $\textsc{QED}$}.
\newblock \emph{\bibinfo{journal}{Phys. Rev. A}} \textbf{\bibinfo{volume}{85}},
  \bibinfo{pages}{041801(R)} (\bibinfo{year}{2012}).

\bibitem{Faraon11}
\bibinfo{author}{Faraon, A.} \emph{et~al.}
\newblock \bibinfo{title}{Integrated quantum optical networks based on quantum
  dots and photonic crystals}.
\newblock \emph{\bibinfo{journal}{New J. Phys.}} \textbf{\bibinfo{volume}{13}},
  \bibinfo{pages}{055025} (\bibinfo{year}{2011}).

\bibitem{Hennessy07}
\bibinfo{author}{Hennessy, K.} \emph{et~al.}
\newblock \bibinfo{title}{Quantum nature of a strongly coupled single quantum
  dot-cavity system}.
\newblock \emph{\bibinfo{journal}{Nature}} \textbf{\bibinfo{volume}{445}},
  \bibinfo{pages}{896--899} (\bibinfo{year}{2007}).

\bibitem{Ishida2013}
\bibinfo{author}{Ishida, N.}, \bibinfo{author}{Byrnes, T.},
  \bibinfo{author}{Nori, F.} \& \bibinfo{author}{Yamamoto, Y.}
\newblock \bibinfo{title}{Photoluminescence of a microcavity quantum dot system
  in the quantum strong-coupling regime}.
\newblock \emph{\bibinfo{journal}{Sci. Rep.}} \textbf{\bibinfo{volume}{3}},
  \bibinfo{pages}{1180} (\bibinfo{year}{2013}).

\bibitem{Xu07}
\bibinfo{author}{Xu, X.~L.} \emph{et~al.}
\newblock \bibinfo{title}{¡°plug and play¡± single-photon sources}.
\newblock \emph{\bibinfo{journal}{Appl. Phys. Lett.}}
  \textbf{\bibinfo{volume}{90}}, \bibinfo{pages}{061103}
  (\bibinfo{year}{2007}).

\bibitem{Xu08}
\bibinfo{author}{Xu, X.~L.} \emph{et~al.}
\newblock \bibinfo{title}{¡°plug and play¡± single photons at 1.3 $\mu$m
  approaching gigahertz operation}.
\newblock \emph{\bibinfo{journal}{Appl. Phys. Lett.}}
  \textbf{\bibinfo{volume}{93}}, \bibinfo{pages}{021124}
  (\bibinfo{year}{2008}).

\bibitem{Chang07}
\bibinfo{author}{Chang, D.~E.}, \bibinfo{author}{S{\o}ensen, A.~S.},
  \bibinfo{author}{Demler, E.~A.} \& \bibinfo{author}{Lukin, M.~D.}
\newblock \bibinfo{title}{A single-photon transistor using nanoscale surface
  plasmons}.
\newblock \emph{\bibinfo{journal}{Nat. Phys.}} \textbf{\bibinfo{volume}{3}},
  \bibinfo{pages}{807--812} (\bibinfo{year}{2007}).

\bibitem{Imamoglu97}
\bibinfo{author}{Imamo\v{g}lu, A.}, \bibinfo{author}{Schmidt, H.},
  \bibinfo{author}{Woods, G.} \& \bibinfo{author}{Deutsch, M.}
\newblock \bibinfo{title}{Strongly interacting photons in a nonlinear cavity}.
\newblock \emph{\bibinfo{journal}{Phys. Rev. Lett.}}
  \textbf{\bibinfo{volume}{79}}, \bibinfo{pages}{1467--1470}
  (\bibinfo{year}{1997}).

\bibitem{Majumdar2012}
\bibinfo{author}{Majumdar, A.}, \bibinfo{author}{Bajcsy, M.},
  \bibinfo{author}{Rundquist, A.} \& \bibinfo{author}{Vu\v{c}kovi\'{c}, J.}
\newblock \bibinfo{title}{Loss-enabled sub-poissonian light generation in a
  bimodal nanocavity}.
\newblock \emph{\bibinfo{journal}{Phys. Rev. Lett.}}
  \textbf{\bibinfo{volume}{108}}, \bibinfo{pages}{183601}
  (\bibinfo{year}{2012}).

\bibitem{Reinhard2012}
\bibinfo{author}{Reinhard, A.} \emph{et~al.}
\newblock \bibinfo{title}{Strongly correlated photons on a chip}.
\newblock \emph{\bibinfo{journal}{Nat. Photon.}} \textbf{\bibinfo{volume}{6}},
  \bibinfo{pages}{93--96} (\bibinfo{year}{2012}).

\bibitem{Brossard10}
\bibinfo{author}{Brossard, F. S.~F.} \emph{et~al.}
\newblock \bibinfo{title}{Strongly coupled single quantum dot in a photonic
  crystal waveguide cavity}.
\newblock \emph{\bibinfo{journal}{Appl. Phys. Lett.}}
  \textbf{\bibinfo{volume}{97}}, \bibinfo{pages}{111101}
  (\bibinfo{year}{2010}).

\bibitem{Zhang14}
\bibinfo{author}{Zhang, W.}, \bibinfo{author}{Yu, Z.~Y.}, \bibinfo{author}{Liu,
  Y.~M.} \& \bibinfo{author}{Peng, Y.~W.}
\newblock \bibinfo{title}{Optimal photon antibunching in a
  quantum-dot-bimodal-cavity system}.
\newblock \emph{\bibinfo{journal}{Phys. Rev. A}} \textbf{\bibinfo{volume}{89}},
  \bibinfo{pages}{043832} (\bibinfo{year}{2014}).

\bibitem{Liew10}
\bibinfo{author}{Liew, T. C.~H.} \& \bibinfo{author}{Savona, V.}
\newblock \bibinfo{title}{Single photons from coupled quantum modes}.
\newblock \emph{\bibinfo{journal}{Phys. Rev. Lett.}}
  \textbf{\bibinfo{volume}{104}}, \bibinfo{pages}{183601}
  (\bibinfo{year}{2010}).

\bibitem{Bamba2011}
\bibinfo{author}{Bamba, M.}, \bibinfo{author}{Imamo\v{g}lu, A.},
  \bibinfo{author}{Carusotto, I.} \& \bibinfo{author}{Ciuti, C.}
\newblock \bibinfo{title}{Origin of strong photon antibunching in weakly
  nonlinear photonic molecules}.
\newblock \emph{\bibinfo{journal}{Phys. Rev. A}} \textbf{\bibinfo{volume}{83}},
  \bibinfo{pages}{021802(R)} (\bibinfo{year}{2011}).

\bibitem{Majumdar121}
\bibinfo{author}{Majumdar, A.}, \bibinfo{author}{Rundquist, A.},
  \bibinfo{author}{Bajcsy, M.} \& \bibinfo{author}{Vu\ifmmode \check{c}\else
  \v{c}\fi{}kovi\ifmmode~\acute{c}\else \'{c}\fi{}, J.}
\newblock \bibinfo{title}{Cavity quantum electrodynamics with a single quantum
  dot coupled to a photonic molecule}.
\newblock \emph{\bibinfo{journal}{Phys. Rev. B}} \textbf{\bibinfo{volume}{86}},
  \bibinfo{pages}{045315} (\bibinfo{year}{2012}).

\bibitem{Brossard13}
\bibinfo{author}{Brossard, F. S.~F.} \emph{et~al.}
\newblock \bibinfo{title}{Confocal microphotoluminescence mapping of coupled
  and detuned states in photonic molecules}.
\newblock \emph{\bibinfo{journal}{Opt. Express.}}
  \textbf{\bibinfo{volume}{21}}, \bibinfo{pages}{16934--16945}
  (\bibinfo{year}{2013}).

\bibitem{Boyd92}
\bibinfo{author}{Boyd, R.}
\newblock \emph{\bibinfo{title}{Nonlinear Optics}}
  (\bibinfo{publisher}{Academic, New York}, \bibinfo{year}{1992}).

\bibitem{Xu04}
\bibinfo{author}{Xu, X.~L.}, \bibinfo{author}{Williams, D.~A.} \&
  \bibinfo{author}{Cleaver, J. R.~A.}
\newblock \bibinfo{title}{Electrically pumped single-photon sources in lateral
  $\textit{ p-i-n}$ junctions}.
\newblock \emph{\bibinfo{journal}{Appl. Phys. Lett.}}
  \textbf{\bibinfo{volume}{85}}, \bibinfo{pages}{3238--3240}
  (\bibinfo{year}{2004}).

\bibitem{Esslinger10}
\bibinfo{author}{Baumann, K.}, \bibinfo{author}{Guerlin, C.},
  \bibinfo{author}{Brennecke, F.} \& \bibinfo{author}{Esslinger, T.}
\newblock \bibinfo{title}{Dicke quantum phase transition with a superfluid gas
  in an optical cavity}.
\newblock \emph{\bibinfo{journal}{Nature}} \textbf{\bibinfo{volume}{464}},
  \bibinfo{pages}{1301--1306} (\bibinfo{year}{2010}).

\bibitem{Deng14}
\bibinfo{author}{Deng, Y.}, \bibinfo{author}{Cheng, J.}, \bibinfo{author}{Jing,
  H.} \& \bibinfo{author}{Yi, S.}
\newblock \bibinfo{title}{Bose-\textsc{E}instein condensates with
  cavity-mediated spin-orbit coupling}.
\newblock \emph{\bibinfo{journal}{Phys. Rev. Lett.}}
  \textbf{\bibinfo{volume}{112}}, \bibinfo{pages}{143007}
  (\bibinfo{year}{2014}).

\bibitem{Mottl}
\bibinfo{author}{Mottl, R.} \emph{et~al.}
\newblock \bibinfo{title}{Roton-type mode softening in a quantum gas with
  cavity-mediated long-range interactions}.
\newblock \emph{\bibinfo{journal}{Science}} \textbf{\bibinfo{volume}{336}},
  \bibinfo{pages}{1570--1573} (\bibinfo{year}{2012}).

\bibitem{Birnbaum05}
\bibinfo{author}{Birnbaum, K.~M.} \emph{et~al.}
\newblock \bibinfo{title}{Photon blockade in an optical cavity with one trapped
  atom}.
\newblock \emph{\bibinfo{journal}{Nature}} \textbf{\bibinfo{volume}{436}},
  \bibinfo{pages}{87--90} (\bibinfo{year}{2005}).

\bibitem{Hennrich05}
\bibinfo{author}{Hennrich, M.}, \bibinfo{author}{Kuhn, A.} \&
  \bibinfo{author}{Rempe, G.}
\newblock \bibinfo{title}{Transition from antibunching to bunching in cavity
  $\textsc{QED}$}.
\newblock \emph{\bibinfo{journal}{Phys. Rev. Lett.}}
  \textbf{\bibinfo{volume}{94}}, \bibinfo{pages}{053604}
  (\bibinfo{year}{2005}).

\bibitem{Majumdar2011}
\bibinfo{author}{Majumdar, A.}, \bibinfo{author}{Kim, E.~D.},
  \bibinfo{author}{Gong, Y.}, \bibinfo{author}{Bajcsy, M.} \&
  \bibinfo{author}{Vu\ifmmode \check{c}\else
  \v{c}\fi{}kovi\ifmmode~\acute{c}\else \'{c}\fi{}, J.}
\newblock \bibinfo{title}{Phonon mediated off-resonant quantum dot-cavity
  coupling under resonant excitation of the quantum dot}.
\newblock \emph{\bibinfo{journal}{Phys. Rev. B}} \textbf{\bibinfo{volume}{84}},
  \bibinfo{pages}{085309} (\bibinfo{year}{2011}).

\bibitem{Majumdar2010}
\bibinfo{author}{Majumdar, A.} \emph{et~al.}
\newblock \bibinfo{title}{Linewidth broadening of a quantum dot coupled to an
  off-resonant cavity}.
\newblock \emph{\bibinfo{journal}{Phys. Rev. B}} \textbf{\bibinfo{volume}{82}},
  \bibinfo{pages}{045306} (\bibinfo{year}{2010}).

\bibitem{Yamaguchi2008}
\bibinfo{author}{Yamaguchi, M.}, \bibinfo{author}{Asano, T.} \&
  \bibinfo{author}{Noda, S.}
\newblock \bibinfo{title}{Photon emission by nanocavity-enhanced quantum
  anti-\textsc{Z}eno effect in solid-state cavity quantum-electrodynamics}.
\newblock \emph{\bibinfo{journal}{Opt. Express.}}
  \textbf{\bibinfo{volume}{16}}, \bibinfo{pages}{18067--18081}
  (\bibinfo{year}{2012}).

\bibitem{Englund2012}
\bibinfo{author}{Englund, D.} \emph{et~al.}
\newblock \bibinfo{title}{Ultrafast photon-photon interaction in a strongly
  coupled quantum dot-cavity system}.
\newblock \emph{\bibinfo{journal}{Phys. Rev. Lett.}}
  \textbf{\bibinfo{volume}{108}}, \bibinfo{pages}{093604}
  (\bibinfo{year}{2012}).

\bibitem{Englund10}
\bibinfo{author}{Englund, D.} \emph{et~al.}
\newblock \bibinfo{title}{Resonant excitation of a quantum dot strongly coupled
  to a photonic crystal nanocavity}.
\newblock \emph{\bibinfo{journal}{Phys. Rev. Lett.}}
  \textbf{\bibinfo{volume}{104}}, \bibinfo{pages}{073904}
  (\bibinfo{year}{2010}).

\bibitem{Auffeves2010}
\bibinfo{author}{Auff\`eves, A.} \emph{et~al.}
\newblock \bibinfo{title}{Controlling the dynamics of a coupled atom-cavity
  system by pure dephasing}.
\newblock \emph{\bibinfo{journal}{Phys. Rev. B}} \textbf{\bibinfo{volume}{81}},
  \bibinfo{pages}{245419} (\bibinfo{year}{2010}).

\bibitem{Majumdar2013}
\bibinfo{author}{Majumdar, A.} \emph{et~al.}
\newblock \bibinfo{title}{Proposed coupling of an electron spin in a
  semiconductor quantum dot to a nanosize optical cavity}.
\newblock \emph{\bibinfo{journal}{Phys. Rev. Lett.}}
  \textbf{\bibinfo{volume}{111}}, \bibinfo{pages}{027402}
  (\bibinfo{year}{2013}).

\bibitem{Flagg2009}
\bibinfo{author}{Flagg, E.} \emph{et~al.}
\newblock \bibinfo{title}{Resonantly driven coherent oscillations in a
  solid-state quantum emitter}.
\newblock \emph{\bibinfo{journal}{Nat. Phys.}} \textbf{\bibinfo{volume}{5}},
  \bibinfo{pages}{203--207} (\bibinfo{year}{2009}).

\bibitem{Ates2009}
\bibinfo{author}{Ates, S.} \emph{et~al.}
\newblock \bibinfo{title}{Non-resonant dot-cavity coupling and its potential
  for resonant single-quantum-dot spectroscopy}.
\newblock \emph{\bibinfo{journal}{Nat. Photon.}} \textbf{\bibinfo{volume}{3}},
  \bibinfo{pages}{724--728} (\bibinfo{year}{2009}).

\bibitem{Tan99}
\bibinfo{author}{Tan, S.~M.}
\newblock \bibinfo{title}{A computational toolbox for quantum and atomic
  optics}.
\newblock \emph{\bibinfo{journal}{J. Opt. B}} \textbf{\bibinfo{volume}{1}},
  \bibinfo{pages}{424--432} (\bibinfo{year}{1999}).

\end{thebibliography}


\newpage
\begin{figure*}
\centering
\epsfig{file=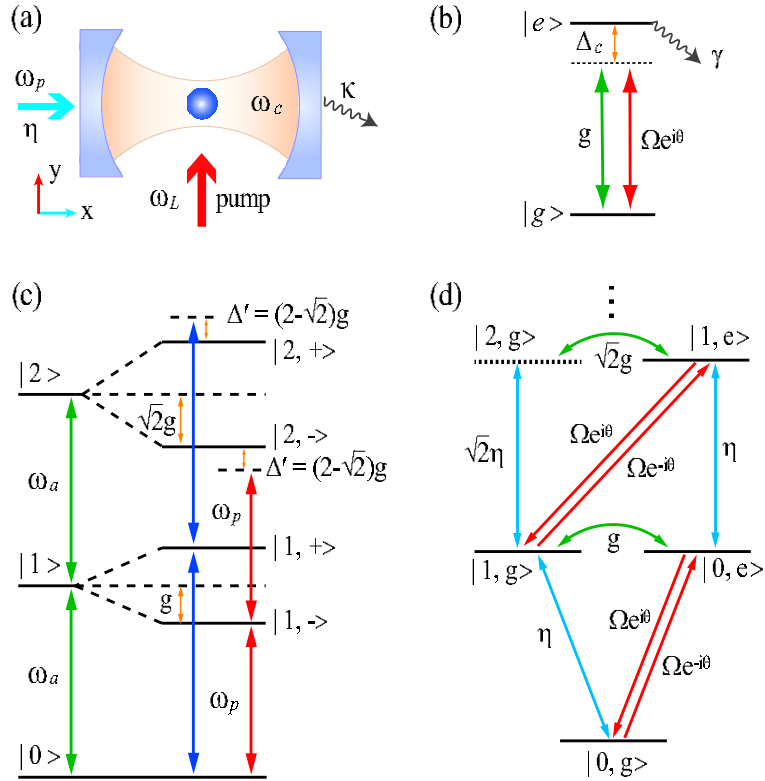,width=10cm,keepaspectratio}
\caption{(a) Scheme for photon blockade of coupling a quantum dot
with a nanocavity. (b) Level diagram for a quantum dot coupled with
the cavity field and the pump field. (c) Energy level diagram of the
dressed states in a coupled quantum dot-cavity system. (d)
Transition paths for the Quantum Interference model.} \label{scheme}
\vspace{5cm}
\end{figure*}

\newpage
\begin{figure*}
\centering
\epsfig{file=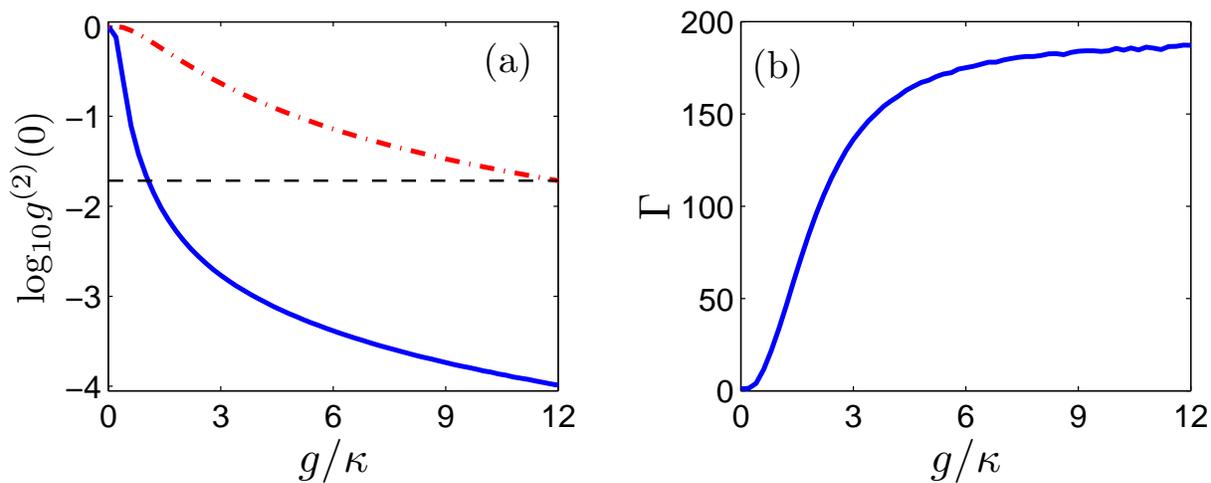,width=16cm,keepaspectratio}
\caption{(a) The minimum second-order correlation function
$g^{(2)}_{\text{JC}}(0)$ (dash-dotted red line) and
$g^{(2)}_{\text{QI}}(0)$ (solid blue line) as a function of the
QD-cavity coupling strength $g$. (b) The ratio $\Gamma =
g^{(2)}_{\text{JC}}(0)/g^{(2)}_{\text{QI}}(0)$ is plotted as a
function of the QD-cavity coupling strength $g$.} \label{g2g}
\vspace{5cm}
\end{figure*}

\newpage
\begin{figure*}
\centering
\epsfig{file=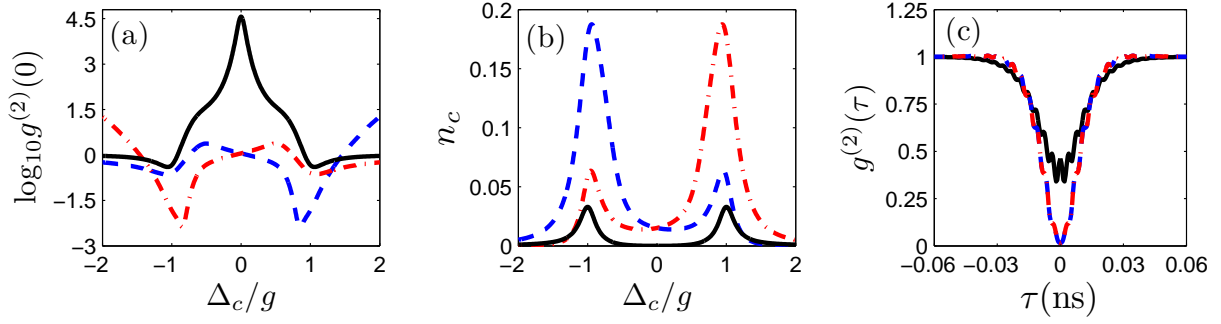,width=16cm,keepaspectratio}
\caption{(a) The second-order correlation function $g^{(2)}(0)$ and
(b) the mean cavity photon number ${n}_c$ as a function of
cavity-light detuning $\Delta_c$. (c) The time-dependent
second-order correlation function $g^{(2)}(\tau)$ of the coupled
system. The solid black lines show the results of JC model with
$\Omega/g=0$. The dashed blue lines and dash-dotted red lines
represent the results with $(\Omega/g, \theta/\pi)=(0.124,0.082)$
and $(\Omega/g, \theta/\pi)=(0.124,\pm1 -0.082)$ in the QI model,
respectively.} \label{photon-number}
\vspace{5cm}
\end{figure*}

\newpage
\begin{figure*}
\centering
\epsfig{file=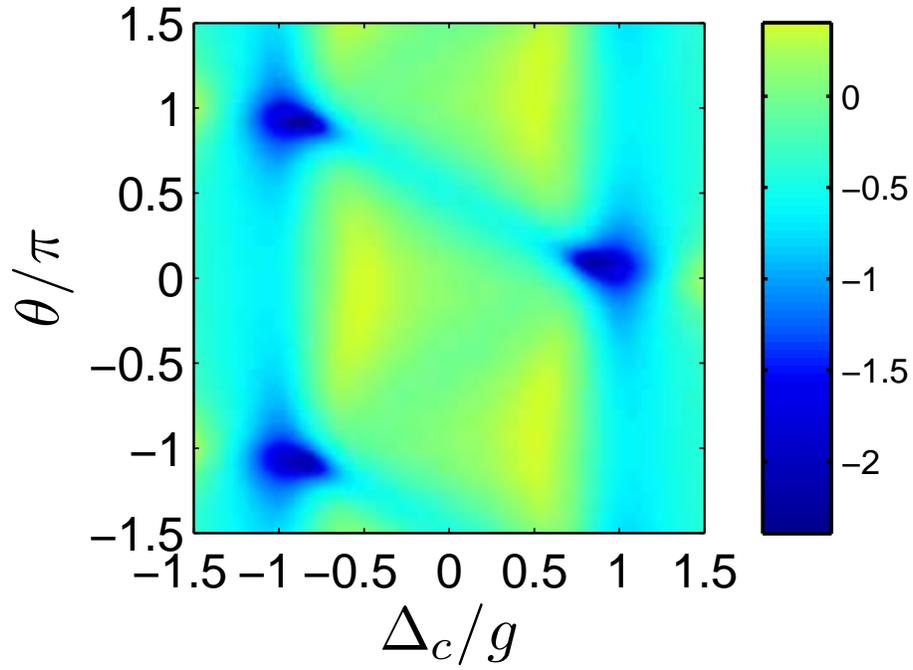,width=12cm,keepaspectratio}
\caption{The second-order correlation function in logarithmic scale (log$_{10}g^{(2)}(0)$)
as a function of cavity-light detuning $\Delta_c$ and relative
dynamic phase $\theta$ for $g=2\kappa$ and $\Omega/g = 0.124$.}\label{theta-Delta}
\vspace{5cm}
\end{figure*}

\newpage
\begin{figure*}
\centering
\epsfig{file=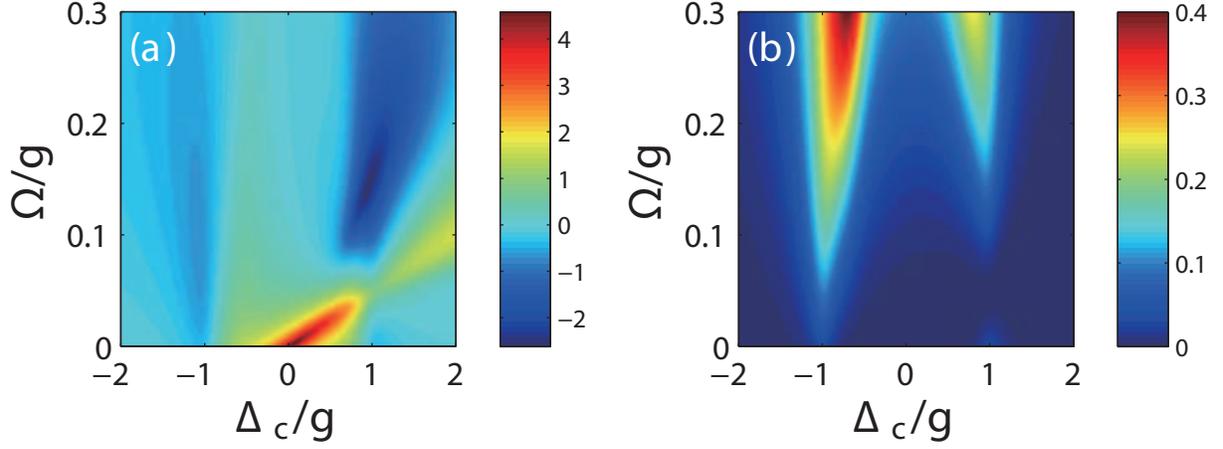,width=16cm,keepaspectratio}
\caption{(a) The second-order correlation function in logarithmic
scale (log10$(g^{(2)}(0))$) and (b) the intracavity photon number
${n}_c$ as a function of cavity-light detuning $\Delta_c$ and Rabi
Rabi coupling strength $\Omega$ for $g=2\kappa$ and
$\theta/\pi=0.082$.} \label{Omega-Delta}
\vspace{5cm}
\end{figure*}

\newpage
\begin{figure*}
\centering
\epsfig{file=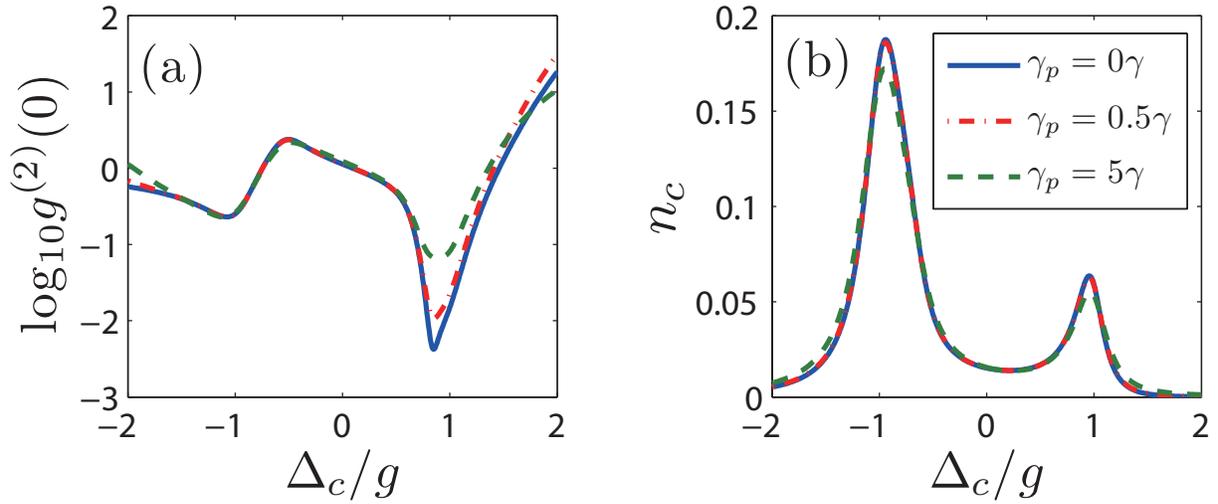,width=16cm,keepaspectratio}
\caption{(a) The second-order correlation function in logarithmic
scale (log10$(g^{(2)}(0))$) and (b) the intracavity photon number
${n}_c$ as a function of cavity-light detuning $\Delta_c$ with
$g=2\kappa$ and $(\Omega/g, \theta/\pi)=(0.124, 0.082)$ for
different pure dephasing $\gamma_d$. The solid blue line, the
dash-dotted red line, and the dashed green line represent the results
with $\gamma_d$ at $0\gamma$, $0.5\gamma$, and
$5\gamma$, respectively.}
\end{figure*}

\end{document}